\documentclass[aps,prl,twocolumn,groupedaddress]{revtex4}

\usepackage{graphics} 
\usepackage{graphicx}
\usepackage{dcolumn}
\usepackage{bm}

\begin{document}

\title{Inter-subband resistance oscillations in crossed electric and magnetic fields}

\author{Scott Dietrich}
\author{Sean Byrnes}
\author{Sergey Vitkalov}
\email[Corresponding author: ]{vitkalov@sci.ccny.cuny.edu}
\affiliation{Physics Department, City College of the City University of New York, New York 10031, USA}
\author{A. V. Goran}
\author {A. A. Bykov}
\altaffiliation{Novosibirsk State Technical University, 630092 Novosibirsk, Russia}
\affiliation{Institute of Semiconductor Physics, 630090 Novosibirsk, Russia}

\date{\today}

\begin{abstract} 
Quantum oscillations of nonlinear resistance are  investigated in response to electric current  and magnetic field applied perpendicular to single GaAs quantum wells with two populated subbands. At small magnetic fields current-induced oscillations  appear as Landau-Zener transitions between Landau levels inside the lowest subband. Period of these oscillations is proportional to the magnetic field.  At high magnetic fields different kind of quantum oscillations emerges with a period,which is independent of the magnetic field. At a fixed current the oscillations are periodic in inverse magnetic field  with a period that is independent of the $dc$ bias. The proposed model considers these oscillations as a result of spatial variations of the energy separation  between two subbands induced by the electric current.

\end{abstract}
  

\maketitle

\section{Introduction}
The magnetotransport phenomena in high-mobility modulation-doped semiconductor structures are commonly studied with only one populated subband ($E_1$), because the electron mobility decreases with filling second subband ($E_2$) due to inter-subband scattering \cite{storm1982}.  The latter also gives rise to  magneto-inter-subband oscillations (MISO) of the dissipative resistance \cite{lead1992}. In electron systems with two populated subbands MISO have maxima in magnetic fields $B$ satisfying the relation \cite{mag,pol,raikh}: $\Delta_{12} = l \cdot \hbar \omega_c$, where $\Delta_{12} = E_2 - E_1$ is the energy separation of the bottoms of the subbands, $\omega_c = eB/m^*$ is cyclotron frequency, $m^*$ is effective electron mass and $l$ is a positive integer. In contrast to Shubnikov de Haas (SdH) oscillations the MIS-oscillations exist at high temperature $kT >\hbar \omega_c$.  An interference of these oscillations with phonon-induced oscillations has been reported\cite{vitkalov2010}.

At small quantizing magnetic fields a finite electric current induces several additional nonlinear phenomena. At low temperatures  small currents decrease considerably the resistance. The dominant  mechanism, inducing the resistance drop, is a peculiar Joule heating (quantal heating), which produces a non-uniform spectral diffusion of electrons over the quantized spectrum. The spectral diffusion is stabilized by inelastic processes ("inelastic" mechanism) \cite{dmitriev2005}. The heating is observed and studied recently \cite{bykov2007,vitkalov2009,bykov2008,gusev2009}.  At higher currents electron transitions between Landau levels occur due to an elastic electron scattering on impurities in the presence of electric field\cite{yang2002,glazman2007}. The transitions increase the resistance, which was  observed in electron systems both with a single occupied subband \cite{bykov2005,zudov2007,dai} and  with multi-subband occupation \cite{bykov2008b,bykovzdrs2010,gusev2011,goran2011}. In the last case an interference of the magneto-inter-subband quantum oscillations (MISO) with the current induced inter-level scattering was reported.
 
Recent investigations of the nonlinear transport in stronger magnetic fields  reveal another kind of current-induced resistance oscillations in electron systems with a single band occupation\cite{vitkalov2012}. These oscillations occur in electric fields that are significantly smaller than the one required for the current-induced Landau-Zener transitions between Landau levels \cite{yang2002}. The period of the current-induced oscillations  is found to be independent of the magnetic field. The oscillations are considered to be a result of spatial variations of the electron filling factor (electron density $\delta n$) with the applied electric field. 

In this paper we report an observation of current-induced resistance oscillations of the dissipative resistance in electron systems with two populated subbands. Two kinds of oscillations are detected. At small magnetic fields we observed resistance oscillations with a period  proportional to the magnetic field. We found that these oscillations are related to the current induced Landau-Zener transitions between Landau levels \cite{yang2002,bykov2008b,gusev2011}. At higher magnetic fields another type of the resistance oscillations emerges with a period that is independent of the magnetic field. In the paper these oscillations are studied at high temperatures at which only MIS-oscillations are present. 

Despite a similarity between the current induced oscillations with the B-independent period, which are found in single subband systems \cite{vitkalov2012} and the oscillations reported in this paper, there is at least one  distinct feature to distinguish the two. Namely the oscillations in the two-subband systems occur at high temperatures $kT \gg \hbar \omega_c$ and, therefore, the total number of the electron states carrying the electric current (inside the energy interval kT) does not oscillate with the Fermi energy (in other words with the  total electron density $n$).   In this regime  the SdH oscillations are damped and  in single subband systems the curren-induced oscillations are absent  \cite{vitkalov2012}.  Thus even if  both kinds of  observed oscillations have a common origin,  the oscillations reported in this paper  are not  directly (simply) related to the spatial variations of the electron density $\delta n$ induced by the electric current.  Another interesting feature is the phase of these oscillations . The oscillations appears to be quasi-periodic with respect to the applied current but with an apparent  $\pi$-phase shift with respect to the zero bias. Below we present our findings and provide an interpretation of the obtained results.

\section{Experimental Setup}

Our samples are high-mobility GaAs quantum wells grown by molecular beam epitaxy on semi-insulating (001) GaAs substrates. The width of the GaAs quantum well is 13 nm. Two AlAs/GaAs type-II superlattices grown on both sides of the well served as barriers, providing a high mobility of 2D electrons inside the well at a high electron density\cite{fried1996}. Two samples were studied with electron density $n_{1,2}$ = 8.09 $\times 10^{15}$ m$^{-2}$ and mobility $\mu_1$= 121 m$^2$/Vs and $\mu_2$= 73 m$^2$/Vs

The studied 2D electron systems are etched in the shape of a Hall bar. The width and the length of the measured part of the samples are $d=$50$\mu m$ and $L=$450$\mu m$. To measure the resistance we  use the four probes method. Direct electric current $I_{dc}$ ($dc$ bias) is applied simultaneously with 12 Hz $ac$ excitation $I_{ac}$ through the same current contacts ($x$-direction). The longitudinal $ac$ ($dc$)  voltage $V^{ac}_{xx}$ ($V^{dc}_{xx}$) is measured between potential contacts displaced 450$\mu m$ along each side of the sample. The Hall voltage $V_H$ is measured between potential contacts displaced 50$\mu m$ across the electric current in $y$-direction.

The current contacts are  separated from the measured area by a distance of 500$\mu m$, which is much greater than the inelastic relaxation length of the 2D electrons $L_{in}=(D \tau_{in})^{1/2} \sim 1-5 $$\mu m$. The longitudinal and Hall voltages were measured simultaneously, using two lockin amplifiers with 10 M$\Omega$ input impedances. The potential contacts provided insignificant contribution to the overall response due to small values of the contact resistance (about 1k$\Omega$) and negligibly small electric current flowing through the contacts.

Measurements were taken at different temperatures and magnetic fields in a He-3 insert inside superconducting solenoid. Samples and a calibrated thermometer were mounted on a cold copper finger in vacuum. Magnetic field was applied perpendicular to the 2D electron layers.

\begin{figure}[t!]
\hskip -0.9 cm
\includegraphics[width=3.5 in]{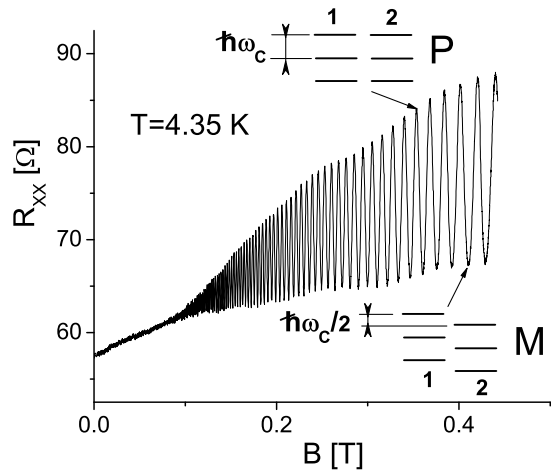}
\caption{Dependence of the resistance $R_{xx}$ on magnetic field with no dc bias applied. Sample N1. }
\label{miso}
\end{figure}

\begin{figure}[t!]
\hskip -0.5 cm
\includegraphics[width=3.5 in]{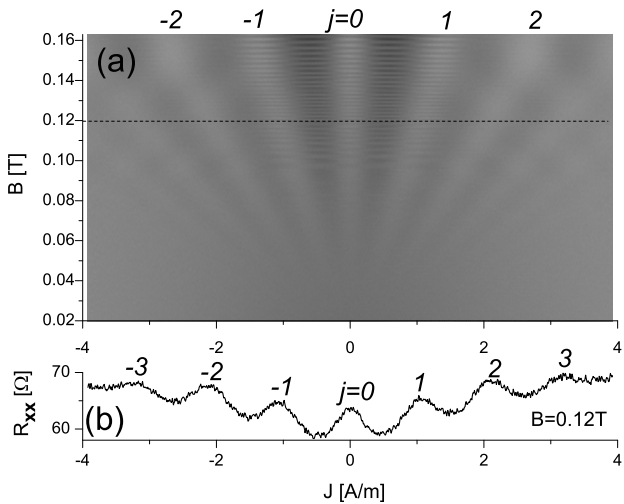}
\caption{a) Dependence of differential resistance $R_{xx}$ on magnetic field and averaged density of electric current $J$; (b) Dependence of the resistance on the current density $J$ at fixed magnetic field as labeled. Index $j=\pm 1,\pm2$... numerates Landau-Zener transitions inside lowest subband, which obey Eq.(\ref{lze}).  T=5.1 K. Sample N1. }
\label{lz}
\end{figure}

\section{Results}

Figure \ref{miso} presents the dependence of the dissipative resistance on the magnetic field at  temperature $T=4.35$ K.  At this temperature $kT >\hbar \omega_c$ and Shubnikov de Haas oscillations are suppressed at $B<$0.5 T. The maximums of the observed magneto-intersubband oscillations (MISO) are due to the enhancement of electron elastic scattering, which occurs when the Landau levels in two subbands are lined up with each other (state P in Fig.\ref{miso}). At this condition elastic electron transitions occur between the subbands, increasing the total electron scattering rate and, thus, the resistance. Minima of the oscillations occur when the Landau levels in one subband are between the levels of another subband. In this condition the elastic electron scattering between subbands is suppressed  (state M in Fig.\ref{miso})\cite{raikh}.

 \begin{figure}[t!]
\includegraphics[width=4in]{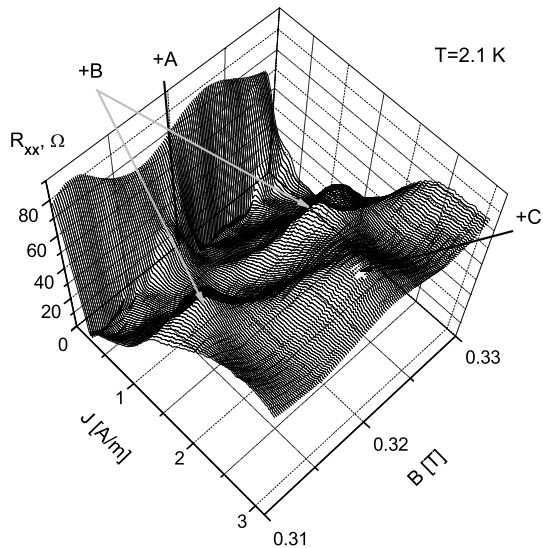}
\caption{Dependence of resistance $R_{xx }$ on magnetic field and current density $J$. Labels +A, +B and  +C indicate different maximums induced by dc bias. T=2.1 K. Sample N1.}
\label{typ}
\end{figure}

Figure \ref{lz}(a) presents differential resistance $R_{xx}$ at different averaged density of the electric current $J=I_{dc}/(d=50\mu$) and small magnetic fields. The differential resistance oscillates with the $dc$ bias. An example of the oscillations is shown in figure \ref{lz}(b) at fixed magnetic field B=0.12 Tesla.  The dependence is a horizontal  cut of the 2D plot and is shown by the dashed line in Fig.\ref{lz}(a). Position of a resistance  maximum $j$  is proportional to the magnetic field and satisfies the following relation:

\begin{equation}
2 eE_j R_c^{(1)} = j \cdot  \hbar \omega_c,
\label{lze}
\end{equation}
where $E_j$ is the electric field (mostly the Hall electric field in the sample) corresponding to the maximum $j$ ,  $R_c^{(1)}$ is the cyclotron radius of electrons in the first subband (the lowest subband) and $j=0,1,2...$  is an integer. Eq.(\ref{lze}) describes Landau-Zener transitions between Landau levels in the first subband\cite{yang2002}. 

 At a higher resolution the data shows oscillations of  the magnitude of the maximums $j= \pm 1$  with the magnetic field at $B>0.1$ (T). The oscillations are periodic in inverse magnetic field and are in-phase  with the intersubband oscillations at zero $dc$ bias ($j=0$). Similar oscillations are observed for the minimum between $j=0$ and $j= \pm 1$ maximums. These oscillations are shifted by phase $\pi$ with respect to the  oscillations of the maximums $j=0,  \pm 1$.  The observed oscillations appear as an interplay between the $dc$ bias  induced Landau-Zener transitions between Landau levels inside the lowest subband and the intersubband transitions, which are periodic in inverse magnetic field $1/B$.  At higher $dc$ biases ($\vert j \vert>1$) the amplitude modulation with the $1/B$ periodicity disappears. In particular no amplitude modulation is found for $j= \pm 2,3$ maximums. 

\begin{figure}[t!]
\includegraphics[width=3.5in]{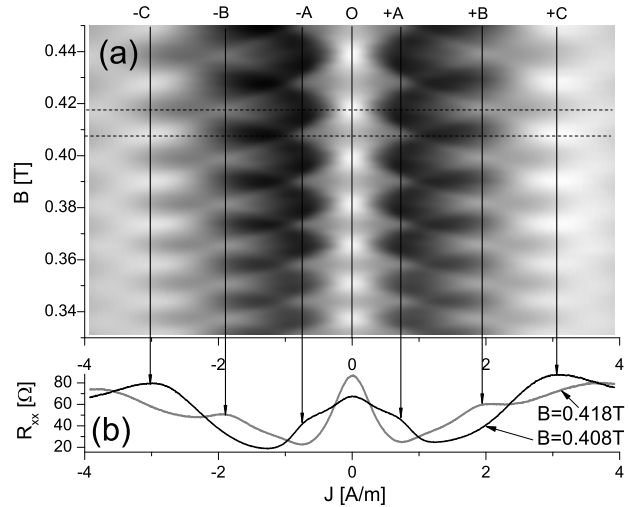}
\caption{(a) Dependence of resistance $R_{xx }$ on magnetic field and current density $J$. indicating strong correlation of features $\pm$A and $\pm$C with MISO minimums and features  $\pm$B with MISO maximums. (b) Dependence of $R_{xx}$ on current density $J$ at magnetic field B=0.418 T corresponding to MISO maximum and  at magnetic field B=0.408 T corresponding to MISO minimum. T= 4.7K. Sample N1.}
\label{evol}
\end{figure}

Figure \ref{typ} presents a typical nonlinear response at a high magnetic field. The response is symmetric with respect to applied $dc$ bias and is shown for the positive bias.  There are several distinct features, which appear with the $dc$ bias. The features are labeled in the figure. Firstly we discuss evolution of the resistance with the dc bias at minimum of a MIS oscillation (state $M$ in fig.\ref{miso}).  When the $dc$ bias is applied the resistance falls down and, then, develops a shoulder labeled by symbol +A. The initial drop of the resistance is mostly due to the quantal heating.   Further increase of the $dc$ current leads to formation of a maximum labeled by symbol +C.  

When the $dc$ bias is applied to state P (see fig.\ref{miso}), corresponding to maximum of a MIS oscillation, the resistance drops much more abruptly and significantly in comparison with the previous case. At low temperatures the resistance drop reaches zero and forms zero resistance state (ZDRS)\cite{vitkalov2007,zudov2010,bykovzdrs2010,gusevzdrs2011}. Further increase of the $dc$ bias leads to the formation of a maximum labeled by symbol +B.

\begin{figure}[t!]
\includegraphics[width=3.5in]{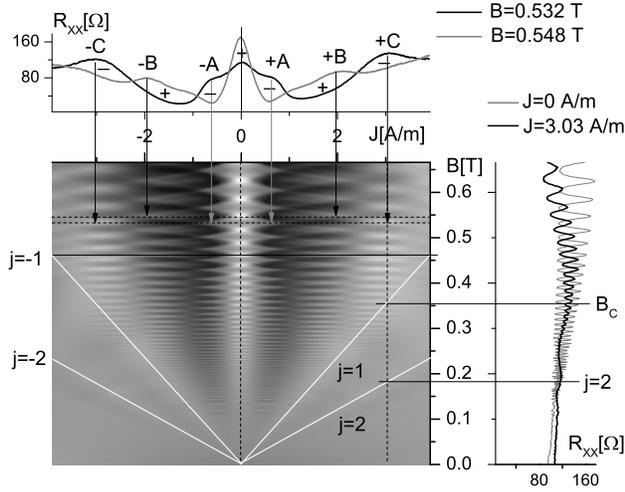}
\caption{Evolution of differential resistance with magnetic field and current density in broad range of magnetic fields. White straight lines indicate Landau-Zener transitions which obey Eq.(\ref{lze}). Upper panel presents horizontal cut  through  MISO maximum at B=0.548T (gray line) and cut through MISO minimum at B=0.532 T (black line).  Sign +(-) indicates regions of current density J, inside which the current induced oscillations have  0(180) degree phase shift  with respect to MIS-oscillations  at J=0A/m. Right panel presents two vertical cuts of the 2D plot   taken at current densities as labeled. Magnetic filed dependence at J=3.03 A/m indicates strong reduction of the resistance oscillations at $B<B_c$ inside the region corresponding to Landau-Zener transitions.  T=5K. Sample N2. }
\label{main}
\end{figure}

\begin{figure}[t!]
\includegraphics[width=3.5in]{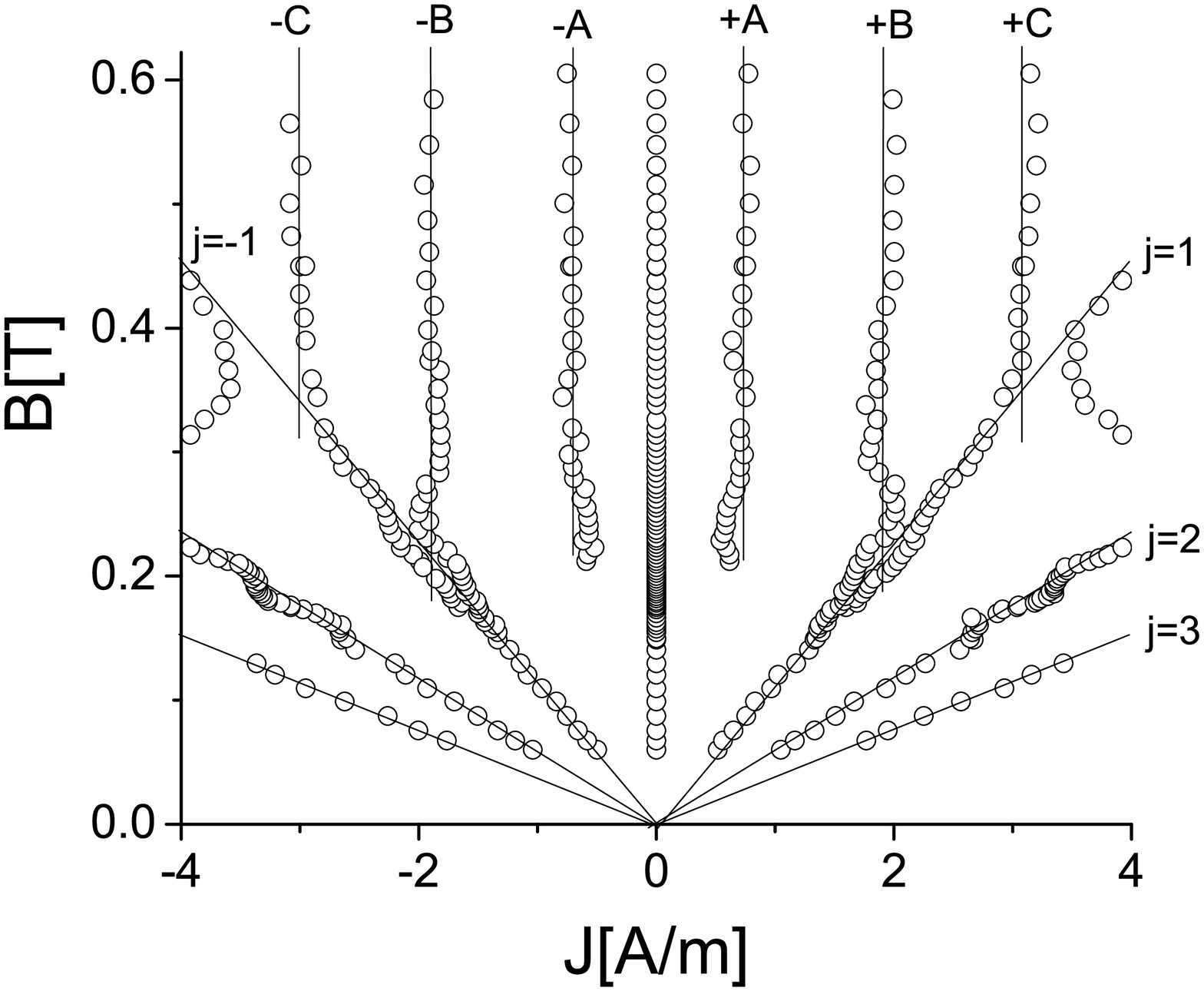}
\caption{ Positions of resistance maximums and different magnetic fields and current density. Two kind of oscillations are observed: in magnetic fields at and below $B_c$, which satisfy  Eq.(\ref{crossover}),  the maximums correspond to Landau-Zener transitions in lowest subband that obey Eq.(\ref{lze}).  Solid straight lines at $j=\pm$1, 2, and 3 represent the equation. At $B>B_c$ the resistance maximums follow the vertical solid lines representing features $\pm$A, $\pm$B and $\pm$C shown on Fig.\ref{typ},\ref{evol},\ref{main}. The crossover between two kind of oscillations occurs at $B=B_c$ presented by line $j=\pm 1$.   Sample N1.   }
\label{main2}
\end{figure}

An evolution of the discussed features with the magnetic field is shown in figure \ref{evol}(a). The figure demonstrates that the positions of all features ($\pm$A, $\pm$B, $\pm$C)    are essentially independent of the magnetic field. Figure \ref{evol}(b)  presents horizontal cuts of the 2D plot through a maximum (B=0.418 T) and a minimum (B=0.408 T) of the inter-subband quantum oscillations.

Figure \ref{main} presents an overall behavior of the quantum oscillations in a broad range of magnetic fields and $dc$ biases. The data was obtained on sample N2.  The figure shows the  crossover  of  the intraband Landau-Zener transitions, obeying Eq.(\ref{lze}),  and  the oscillations marked as  $\pm$A, $\pm$B, $\pm$C, which have the MISO periodicity.  The apparent  crossover occurs near the Landau-Zener transition corresponding to $j=\pm 1$. Namely the oscillations with 1/B MISO periodicity occurs at magnetic fieds $B_c$ corresponding to   

\begin{equation}
\hbar \omega_c \ge 2 eE_1 R_c^{(1)}.   
\label{crossover}
\end{equation}

At smaller magnetic fields $B<B_c$ the oscillations are significantly reduced.  Two vertical cuts of the 2D plot taken at different currents   are shown in the right panel of Fig. \ref{main}. The curve taken at J=3.03 A/m shows  the strong reduction of the oscillations at $B<B_c$ in a comparison with the MISO  at J=0A.  Thus the main intraband Landau-Zener transition ($j=\pm 1$) forms a boundary below which the current induced oscillations with 1/B intersubband periodicity are strongly damped.

The upper panel of Fig. \ref{main} shows two horizontal  cuts of the 2D plot.  Black solid line presents dependence of the resistance $R_{xx}$ on $dc$ bias taken at B=0.532(T) corresponding to a mimumum of MISO. The grey line presents the dependence taken at $B=0.548$ (T) corresponding to a MISO maximum. The two curves intersect at 8 points. These intersections marks the regions at which the oscillations with MISO periodicity changes their phase by $\pi$.  At the intersections the oscillations  are nearly vanished.   Sign "+" indicates the region between two intersections in which  the oscillations are in-phase with the MISO, whereas sign "-" indicates the regions in which the oscillations are shifted by phase $\pi$ with respect to the MISO.  

Figure \ref{main2} presents an accurate position of the resistance maximums  with 1/B periodicity at different currents and magnetic fields for sample N1. The figure indicates clearly that at $B=B_c$ ($j=\pm 1$)  the resistance maximums follow the main Landau-Zener transition $j=\pm 1$ whereas at $B>B_c$ the maximums are nearly independent on magnetic field (features  $\pm$A, $\pm$B, $\pm$C).  The solid  lines $j=\pm 1$ mark the boundary between the two kinds of oscillations. The lines obey eq.(\ref{lze}) at $j=\pm 1$ with the cyclotron radius $R_c^1$ corresponding to the lowest subband. The complete theory of the current induced oscillations of the resistance of 2D electron system with two populated subbands is not available in a general case. The case of a bilayer electron system with two closely spaced and almost equally populated electronic subbands has been studied recently\cite{bykov2008b,gusev2011}. These results are in qualitative agreement with the present data at small magnetic fields $B<B_c$.

At high magnetic fields $B>B_c$ figures \ref{main} and \ref{main2} presents new kind of the current induced quantum oscillations. A striking feature of these oscillations is the independence of the position of these oscillations on magnetic field. An interesting property of these oscillations is the region in which the oscillations occur. Figures \ref{main}, \ref{main2} show that these oscillations  start at the line corresponding to Landau-Zener transitions at $j=\pm 1$ in the lowest subband and propagate to higher magnetic fields.  Another interesting property is an apparent quasi-periodicity of the oscillations with applied current. Namely the features  $\pm$A, $\pm$B, $\pm$C are displaced by about the same value  of the electric current density from each other: $\delta J \sim $1.27 (A/m).  The phase of the oscillations is shifted by $\pi$ with respect to zero dc bias. It seems strange that the  MIS-oscillations (J=0 (A/m) are not a  part of this periodic set.  

\begin{figure}[t!]
\includegraphics[width=3.5in]{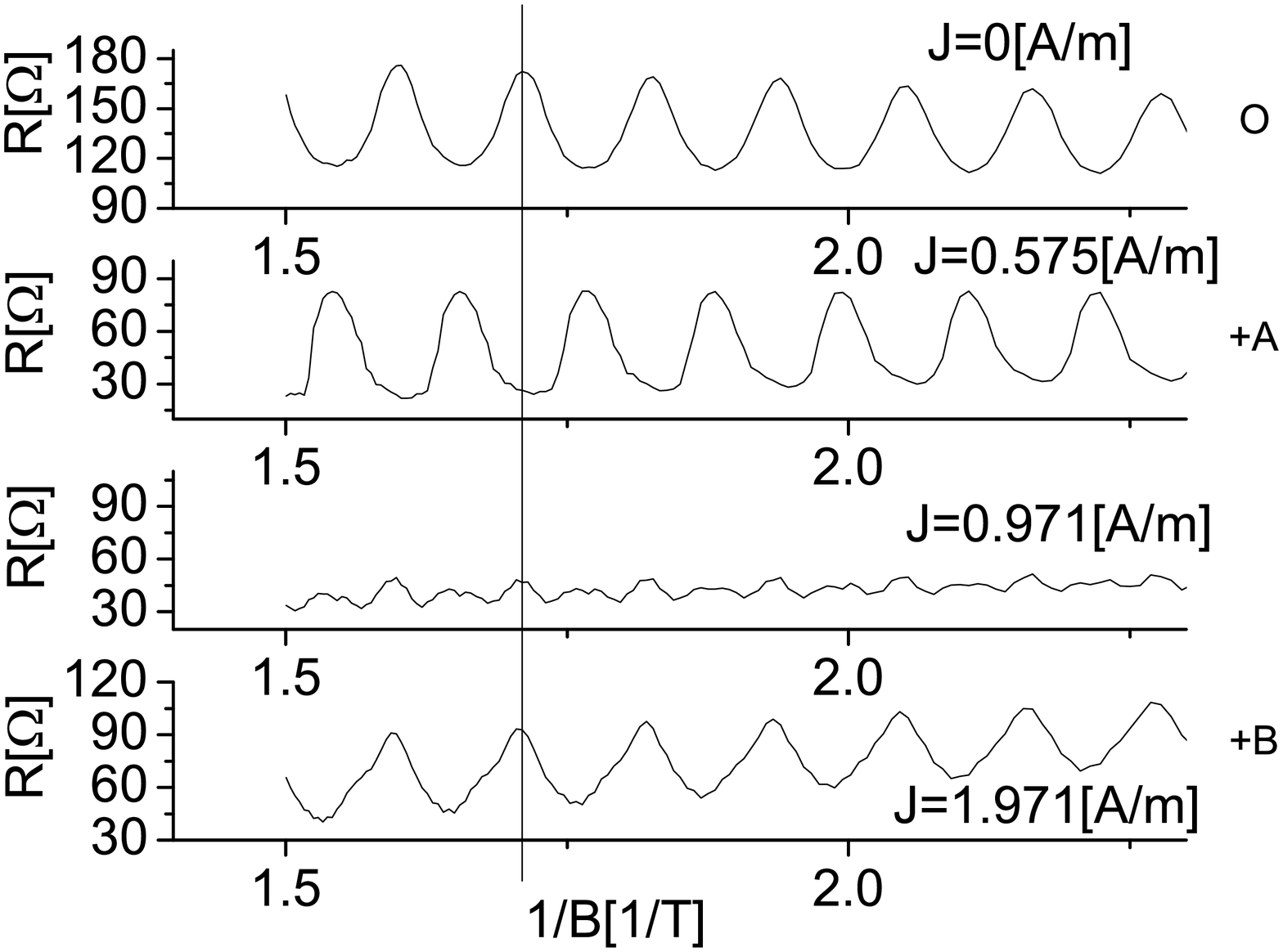}
\caption{Dependence of resistance on inverse magnetic field at different dc biases as labeled.  T=5 K. Sample N2.}
\label{fft}
\end{figure}

Figure \ref{fft} demonstrates the $1/B$ periodicity and the phase of the current induced  oscillations at different  $dc$ biases as labeled. The figure indicates that  oscillations at J=1.97 (A/m) ( B+ feature) are in phase with MISO, whereas oscillations at J=0.575 A/m (A+ feature) are shifted by $\pi$ with respect to MISO.  Figure \ref{fft} shows also the strong reduction of the oscillations at J=0.971 A/m. At this current the oscillations change phase by $\pi$. The current corresponds to the intersection of two curves shown in upper panel of Fig.\ref{main}. 

The 1/B periodicity of the oscillations and the magnetic field independence of the electric current $I_{dc}$,  inducing the  oscillations at  $B>B_c$,  indicates a similarity of  these quantum oscillations with the current induced quantum oscillations reported recently in Ref.\cite{vitkalov2012}.  Below we consider a model, which is, in many respects, analogous  to one described in Ref.\cite{vitkalov2012}. The model  reproduces  the main properties of the observed quantum oscillations.

\section{Model and Discussion} 
 
Current induced quantum oscillations with 1/B periodicity were recently observed in 2D electron systems with  a single occupied subband \cite{vitkalov2012}. The oscillations occur in a strong magnetic field at which Shubnikov de Haas  oscillations (SdH)\cite{shoenberg1984} are well developed. With respect to the electric current the oscillations are periodic with a period that is  independent on the magnetic field.  The proposed model considers the oscillations as  result of a variation of the electron filling factor with the dc bias. In contrast to SdH oscillations, the variation appears across the sample and is related to a spatial change of the electron density $\delta n$.  If the change $\delta n$ is comparable with the number of electron states in a Landau level $n_0=m/(\pi \hbar^2) \cdot \hbar \omega_c$, then one should expect a variation of the electron resistivity.  The spatial variation of the resistivity leads to  oscillations of the sample resistance\cite{vitkalov2012}. 

MIS-oscillations occur due to a periodic enhancement of the inter-subband scattering, when Landau levels in two subbands are line up as shown in Fig.\ref{miso}.    MISO have maxima in magnetic fields $B$ satisfying the relation \cite{mag,pol,raikh}: $\Delta_{12} = i\hbar \omega_c$, where $\Delta_{12} = E_2 - E_1$ is the energy separation of the bottoms of the subbands and $i$ is an integer. In contrast to SdH oscillations the MIS-oscillations exist at high temperature $kT >\hbar \omega_c$ and are insensitive to variations of the Fermi energy and/or electron density $n$ for non-interacting  2D carriers.  

For  interacting electron systems the  situation is different.  Recent direct experiment indicates that gap $E_0$ between  conducting  and valence bands of 2D electron systems formed in GaAs quantum wells depends considerably on the electron density $n$ \cite{ashk2011}. This observation opens a  way to consider the dependence of the energy separation between two subbands $\Delta_{12}$ on the electron density  as  a mechanism  leading to the current induced quantum oscillations in magnetic fields $B>B_c$. Indeed  the experiment Ref.\cite{ashk2011} demonstrated  about one percent change of the gap $E_0$ at a Hall voltage $V_H$=75 (mV) in magnetic field B=0.3 (T). The Hall voltage is comparable with the  one observed in our experiment: $V_H \approx $ 50 mV at B=0.35 (T) and J=4 (A/m). At B=0.35 (T) the phase of the MISO $2\pi \Delta_{12}/\hbar \omega_c \approx 2\pi \cdot 30$  requires about 3 percent change of the inter-subband energy separation $\Delta_{12}$ to make an additional MIS-oscillation cycle.  The comparison indicates a feasibility of the proposed mechanism,   taking into account that in our samples the GaAs quantum well is sandwiched between  conducting layers, which  enhance  significantly the electron screening and, therefore, the variations of the electron density $\delta n$ with the dc bias \cite{vitkalov2012}.

In the model described below  we assume that the dc bias-induced variation of the electron density $\delta n(r)$ changes the energy separation $\Delta_{12}(n)$ between two subbands across samples. Since relative variations of the electron density  is small $\delta n/n \ll 1$, we will consider  only the linear term of the dependence $\Delta_{12}(n)$: 

\begin{equation}
\Delta_{12}(n) =\Delta_0+\gamma \delta n(r),
\label{ndep}
\end{equation}
where $\Delta_0$ is the energy separation at zero dc bias and the parameter $\gamma$ is a constant.  The following  consideration is qualitatively similar to the model described in detail in Ref.\cite{vitkalov2012}.  Below we describe the main parts of the model omitting some details.    

The conducting 2D electron system in the GaAs quantum well is sandwiched between two layers of AlAs/GaAs superlattices (SL) of the second kind \cite{fried1996}.   The parameters of the superlattices are adjusted to set the system close to a metal-insulator transition. At this condition the barely-conducting SL layers efficiently screen  electric charges but do not contribute considerably to the overall conductivity of the structure.  Electric contacts connect the GaAs and the SL layers. Thus the system is considered as a set of parallel conductors. At zero magnetic field the distribution of the electric potential driving the current  is the same in all layers due to the same shape of the conductors. That is to say at B=0 the potential difference between different layers is absent. In the poorly conducting SL layers the electric current is  several order of magnitude smaller than the one in the highly conducting GaAs quantum well.

\begin{figure}[t!]
\vskip -2 cm \hskip -2 cm 
\includegraphics[width=3.7 in]{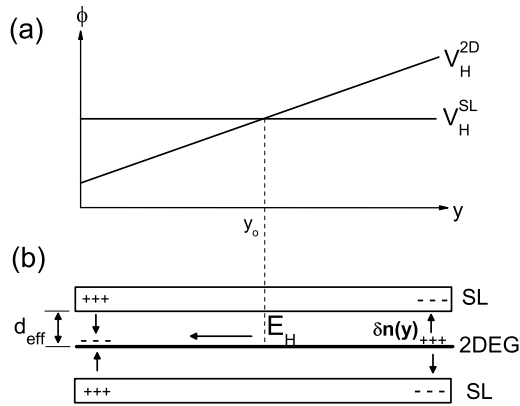}
\caption{ Dependence of the electric potential on position $y$ in the direction perpendicular to the electric current in strong magnetic field. Line $V_H^{2D}$ describes the potential in GaAs quantum well, in which strong Hall effect is developed. Line $V_H^{SL}$ describes the potential in the highly resistive superlattice layer, in which the Hall voltage is negligibly small due to the negligibly small current in the layer.  }
\label{sample2}
\end{figure}

The layers have a different distribution of the electric potential in a strong magnetic field, at which $\omega_c \tau_{tr}^{2D} \gg 1$ and $\omega_c \tau_{tr}^{SL} \ll 1$, where $\tau_{tr}^{2D}$ and $\tau_{tr}^{SL}$ are transport times in the GaAs and in the SL layers.   At $\omega \tau_{tr}^{2D} \gg 1$ the electric field in the GaAs layer is almost perpendicular to the current due to the strong Hall effect.  In contrast the very small electric current in the SL layer induces a Hall voltage, which is negligible.  The Hall voltages are shown in Fig.\ref{sample2} (a) for small currents ( linear response). Figure \ref{sample2}(b) presents distribution of electric charges in the structure. Electric charges are accumulated near the edges of the 2D highly conducting GaAs layer, inducing the Hall electric field $E_H$. The charges are partially screened by charges accumulated in the conducting SL layers. 

Due to the small Hall voltage $E_H^{SL}$ and the absence of the electric current across the system the change of the electric potential $\phi^{SL}(y)$ in the SL  layer is negligibly small. Below we consider the potential $\phi^{SL}$ as a constant. Due to a finite screening length $\lambda_s$ in the SL layer the charge accumulation occurs at a distance $d \sim \lambda_s$. Below we approximate the charge distribution by a charged capacitor with an effective distance $d_{eff}$ between conducting plates.

The proposed model considers a long 2D Hall bar with a width $L_y$\cite{shashkin1986,dyakonov1990}. Electric current is in $x$-direction and the Hall electric field is in $y$-direction. In  a long conductor the electric field $\vec E=(E_x, E_y)$ is independent on $x$, due to the uniformity of the system in $x$ direction:  
\begin{equation}
\frac{\partial E_x}{\partial x} = \frac{\partial E_y}{\partial x}=0
\label{long}
\end{equation}

For a steady current Maxwell equations yield:

\begin{equation}
\frac{\partial E_x}{\partial y} = \frac{\partial E_y}{\partial x}
\label{max}
\end{equation}

Eq.(\ref{long}) and eq.({\ref{max}}) indicate, that the $x$ component of the  electric field is the same at any location: $E_x=E=$const. 

Boundary conditions and the continuity equation require that the density of the electric current in $y$ direction is zero: $J_y=0$ and therefore,

\begin{equation}
E_x=\rho_{xx}J_x \hskip 0.5 cm E_y=\rho_{yx} J_x 
\label{E}
\end{equation} 
where $\rho_{xx}$ and $\rho_{yx}$ are longitudinal and Hall components of the resistivity tensor \cite{ziman}.  We approximate the MIS-oscillations of the resistivity by a simple expression \cite{raikh}:

\begin{equation}
\rho_{xx}(n(y))=\rho_D[1+\alpha \cdot cos(\frac{2\pi \Delta_{12}}{\hbar \omega_c}) ]
\label{miso_t}
\end{equation}   
where $\rho_D$ is Drude resistivity,  and $\alpha$ describes the amplitude of the quantum oscillations.

An electrostatic evaluation of the voltage between conducting layers, shown in Fig.\ref{sample2}(b), yields:

\begin{equation}
\phi^{2D}(y)=\phi^{SL}+\frac{e\delta n(y)d_{eff}}{2\epsilon \epsilon_0}
\label{voltage}
\end{equation}
where $\phi^{2D}$ and $\phi^{SL}$ are electric potentials of the GaAs (2DEG) and superlattice (SL) layers, and $\epsilon$ is permittivity of the SL layer. 
Expressing the electron density $\delta n$ in terms of electric potential $\phi^{2D}$ from Eq.(\ref{voltage}) and  substituting the relation into eq.(\ref{ndep}) and then into eq.(\ref{miso_t}) one can find dependence of the resistitivity on the electric potential: $\rho_{xx}(\phi^{2D})$. 

The relation $E_y=-d\phi^{2D}/dy$ together with  eq.(\ref{E}) yields:
\begin{equation}
-\frac{d\phi^{2D}}{dy} \rho_{xx}(\phi^{2D})=\rho_{yx}E
\label{main_eq}
\end{equation}          

Separation of the variables $\phi^{2D}$ and $y$ and subsequent integration of eq.{\ref{main_eq} between two sides of the 2D conductor ($y$-direction) with corresponding electric potentials $\phi_1$ and $\phi_2$ yield the following result: 
\begin{equation}
\matrix{ \rho_{D}(\phi_2-\phi_1+\frac{2\alpha}{\beta}[sin[\frac{\beta}{2} (\phi_2-\phi_1)] \cr 
\times cos[\frac{\beta}{2}(\phi_2+\phi_1)+\theta_0]])=  \rho_{xy}EL_y \cr 
\cr
\beta=4\pi \epsilon_0 \epsilon \gamma/(e d_{eff} n_0), \cr
\theta_0=2\pi \Delta_0/\hbar \omega_c-\beta \phi^{SL},\cr
}
\label{pr1}
\end{equation}
where $L_y$ is a width of the sample. Taking into account that longitudinal voltage is $V_{xx}=EL_x$, where $L_x$ is a distance between the potential contacts, and the Hall voltage $V_H=\phi_2-\phi_1= -\int E_y dy= -\rho_{yx}I$ (see eq.\ref{E}), the following relation is obtained:

\begin{equation}
V_{xx}=R_{D}(I-\frac{2\alpha}{\beta \rho_{xy}}[sin(\frac{\beta \rho_{xy}I}{2})\cdot cos[\frac{\beta}{2}(\phi_2+\phi_1)+\theta_0]])
\label{final}
\end{equation}
, where $R_D=L_x\rho_D/L_y$ is Drude resistance. 

Eq.\ref{final} is  simplified further for two cases corresponding to a minimum and a maximum of MIS-oscillations. In these cases the voltage  $\phi^{2D}(\delta y)-\phi^{SL}$ is expected to be an asymmetric function of the relative position $\delta y=y-y_0$ with respect to the center of the sample $y_0$  (as shown in fig. \ref{sample2}) and, thus,  $\phi_1-\phi^{SL} =-(\phi_2-\phi^{SL})$ and the argument of the cosine in eq.\ref{final} becomes to be independent on the electric current. In these cases the differential resistance $r_{xx}=dV_{xx}/dI$ is found to be

\begin{equation}
r_{xx}=R_D[1+\alpha \cdot cos(2\pi\frac{I}{I_0})\cdot cos(\frac{2 \pi \Delta_0}{\hbar \omega_c})],
\label{final2}
\end{equation}    
where $I_0=e^3 \hbar d_{eff} n/\epsilon \epsilon_0 m \gamma$.

Eq. \ref{final2} demonstrates oscillations of the differential resistance with the electric current. The period of the oscillations $I_0$ does not depend on the magnetic field  in accordance with the experiment.  The amplitude of the MIS-oscillations is strongly modulated by the dc bias. In particular at $I=I_0/4$ the amplitude is zero. At this node the 1/B periodic oscillations change phase by $\pi$.  The strong amplitude modulation with the dc bias and the  $\pi$ phase shift  at a node  agree with the experiment. 

 Following  from   Eq.(\ref{final2}) positions of the nodes and anti-nodes of the oscillations with respect to the current $I_{dc}$  do not agree with the experiment. In accordance with  Eq.(\ref{final2})  the nodes occurs at 

\begin{equation}
\matrix{J_k=\frac { I_0}{4} \cdot k, \cr
k=2i-1; i=1,2,3...,}
\label{index}
\end{equation}    
where $k$ is a node index. Upper panel of Fig.\ref{main} shows nodes at 0.22, 0.93, 2.41 and 3.91 A/m. Thus the relative positions of the nodes observed in the experiment do not follow the node positions (or index k) in Eq.(\ref{index}). Below we show that the disagreement is reduced significantly taking into account the Joule's heating.

 The discussed above model does not take into account the dc heating of the 2D electrons.  The Joule heating in systems with a discrete spectrum (quantal heating) has a peculiar form providing strong impact on the electron transport \cite{vitkalov2009}. In electron systems with two subbands occupied the quantal heating inverts the MIS-oscillations \cite{bykov2008,gusev2009}. A quantitative account of the  heating  will not be done in this paper. Instead we will use an analytical approximation of the  heating which is valid for two subbands with  equal electron population. As shown below the approach yields the positions of the nodes which agree with the experiment.

The expression for the resistivity  of 2D electron systems with two equally populated subbands in crossed electric and quantizing magnetic field reads \cite{gusev2009}
 \begin{equation}
\matrix{ \rho_{xx}=\rho_D[1+exp(-\frac{2\pi}{\omega_c \tau_q}) \frac{1-3Q}{1+Q}(1+cos(\frac{2 \pi \Delta_{12}}{\hbar \omega_c})] \cr
Q=\frac{2\pi^3 J^2}{e^2n\omega_c^2}\cdot \frac{\tau_{in}}{\tau_{tr}},
}
\label{heat}
\end{equation}     
where $\tau_q$ is quantum scattering time, $\tau_{in}$ and $\tau_{tr}$ are inelastic and transport scattering times. 
To account the heating we replace Eq.(\ref{miso_t}) by Eq.(\ref{heat})  and evaluate differential Eq.(\ref{main_eq}) numerically with fitting  parameters approximating the experimental data. Due to a quite rough approximation of the heating,  the fitting parameters may deviate significantly from actual physical values.  To find the fitting parameter corresponding to the  inelastic scattering time we use that the second term of  Eq.(\ref{heat}) is zero at Q=1/3 \cite{gusev2009}. Assuming that at a small dc bias and low temperatures the quantal heating dominates \cite{vitkalov2009,gusev2009}, we related the first node shown in Fig.\ref{main} at J=0.22 A/m to the condition Q=1/3.   This yields $\tau_{in}=1.8$ ns at  B=0.53T.  Using this value we solved Eq.(\ref{main_eq}) numerically. The result is shown in Fig.\ref{sim} (a). At small dc bias $J \approx $0.17 A/m the figure demonstrates the oscillation node, induced by the heating with a small contribution from the variation of the band separation $\Delta_{12}$. Other nodes occur at considerably higher dc biases and are shifted   with respect to the nodes shown in Fig.\ref{sim}(b)), which  obtained by  the numerical evaluation, ignoring the quantal heating (Q=0). 

At $Q>1/3$ the heating not only shift the nodes but also inverts of the oscillations induced by variations of the band separation. Namely  shown in Fig.\ref{sim}(a) the maximum at J=1.75 A/m  is a result of the dc bias induced evolution of the MISO maximum at J=0A/m. Without the heating the MISO maximum evolves into a  minimum at J=1.65 A/m shown in Fig.\ref{sim}(b).  Thus the heating inverts minimums to maximums and visa versa.  The inversion is directly related to the sign change of  the second term in Eq.(\ref{heat}) at Q=1/3.  

\begin{figure}[t!]
\vskip 0 cm \hskip 0 cm 
\includegraphics[width=3.6 in]{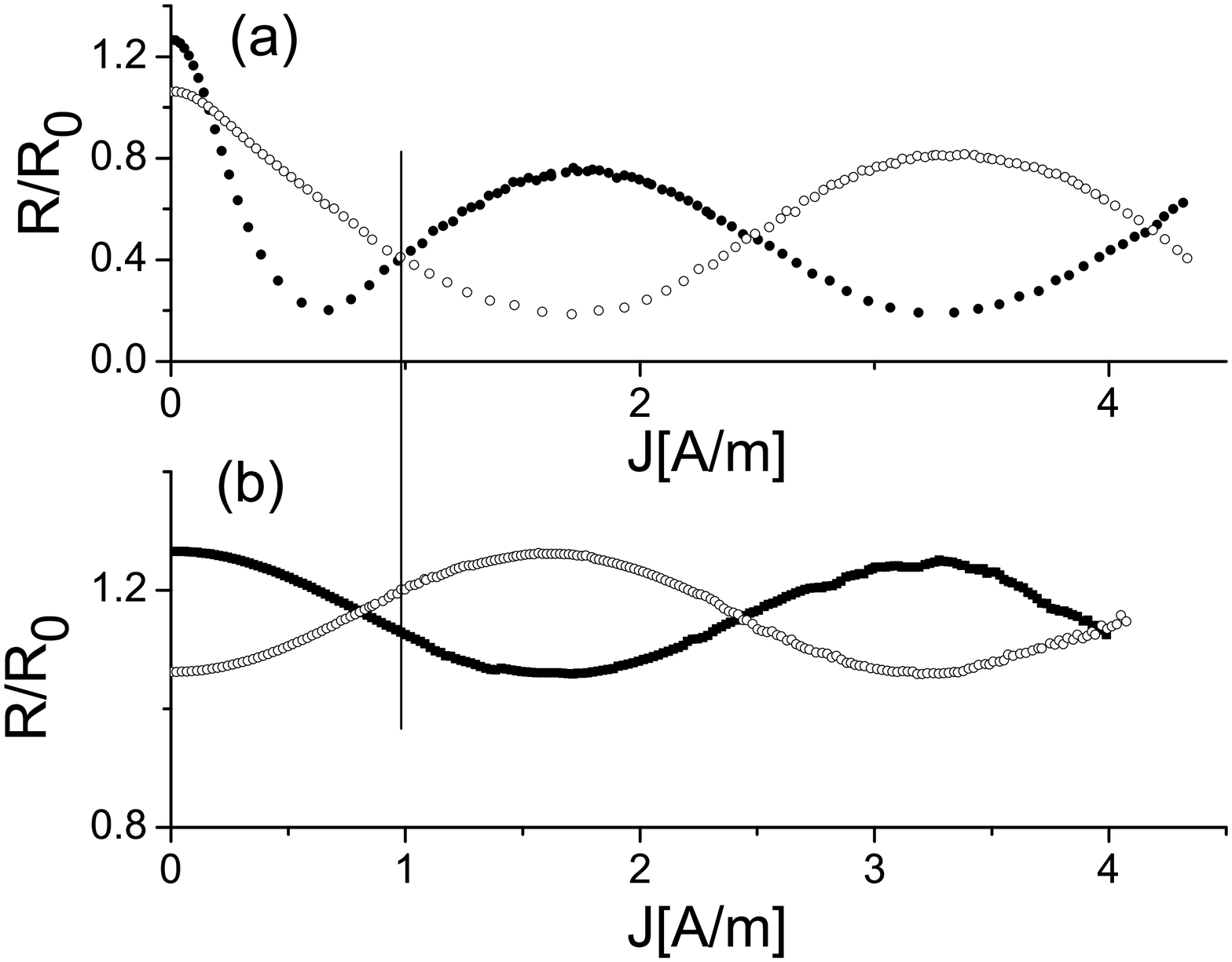}
\caption{ (a) Numerical simulation of the dependence of differential resistance on dc bias at B=0.53 T. Fitting parameters used in the numerical simulation: $\tau_{in}$=1.8 ns,  $\tau_q$=2.5 ps and $\tau_{tr}$=45 ps; electron density n=8.09$\cdot$10$^{15}$ 1/m$^2$; effective screening length $d_{eff}$=30 nm; parameter $\gamma$=1$\cdot$10$^{-37}$ Jm$^2$ (see Eq.(\ref{ndep})). (b) Numerical simulation of the dependence of differential resistance on dc bias  with the same fitting parameters  as in (a) but without dc heating: $\tau_{in}$=0 ns (Q=0). Filled (open) circles present evolution of a MISO maximum (minimum) with the dc bias}
\label{sim}
\end{figure}

The heating and the variation of the band separation affect differently the maximums and minimums of MIS-oscillations. Conversely, quantal heating  decreases the resistance at any magnetic field. A variation of the resistance, induced by the  change of the band separation, depends on the magnetic field. At a maximum (state P in Fig.(\ref{miso})), a variation of $\Delta_{12}$ destroys the level alignment decreasing the inter-band scattering and, thus, the resistance.    At a minimum (state M in Fig.(\ref{miso}), a variation of $\Delta_{12}$ improves  the level alignment and increases the inter-band scattering and the resistance.  Thus at a MISO maximum both the heating and the variations of the band separation decreases the resistance whereas at a MISO minimum two mechanisms work against  each other. In result the drop of the resistance at a MISO maximum is considerably stronger the one at a MISO minimum.  In fact the shoulder (feature +A  in Fig.\ref{typ}) is a result of the competition between two mechanisms at a MISO minimum whereas  ZDRS states, developed from MISO maximums, is a strong indication of the joint decrease of the resistance due to both mechanisms. The behavior is reproduced  in the proposed model. Indeed Fig.\ref{sim}(a) shows that  the initial drop of the MISO maximum is considerably stronger than the decrease of the MISO minimum with the dc bias.   
 
\begin{figure}[t!]
\vskip 0 cm \hskip 0 cm 
\includegraphics[width=3.6 in]{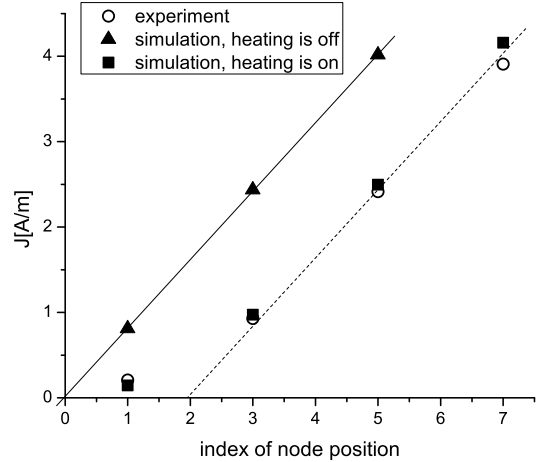}
\caption{ Position of  nodes of dc bias induced oscillations shown in Fig.\ref{sim} and Fig.\ref{main} at different  node index k. Filled triangles present nodes,  which are obtained numerically without heating and  obey Eq.(\ref{index}) (solid line). Account of the heating (filled squares) improves significantly agreement with  the experiment (open circles).  Dashed line is a shift of the solid line to the right by two units  (see text for detail).  }
\label{nodes}
\end{figure}

Figure \ref{nodes} presents  a comparison of  the positions of oscillation nodes, obtained in the model,  with  the experiment. For the purpose of a comparison,  the node positions are plotted versus  the index k, which is defined in  Eq.(\ref{index}).    Without the heating nodes of oscillations obey Eq.{\ref{index}.  Filled triangles demonstrate this behavior. When the heating is on ( filled squares) the first node ( k=1) is due mostly to  the heating. The following nodes ( k=3,5, and 7) are due mostly to the variation of the band separation. As shown in the figure the positions of the nodes correlate well with the experimental values (open circles)  taken from the  upper panel of Fig.\ref{main}. 

The quantal heating produces an additional node of the dc bias induced oscillations. It changes the systematic placement of the node positions described by Eq.(\ref{index}). In the case of a  strong quantal heating (as in  Fig.\ref{sim}) the additional node occurs at the very beginning   of the resistance evolution.  Expected from Eq.(\ref{index}) node counting  can be largely restored by a reduction of the node index by two, which is the difference between consecutive indexes k in Eq.(\ref{index}). The corresponding transformation is shown in Fig.\ref{nodes}: the dashed line is the shift by two units to the right of  the solid line representing index k in Eq.(\ref{index}).

\section{Conclusion}

Quantum oscillations of nonlinear resistance, which occur in response to electric current  and magnetic field applied perpendicular to GaAs quantum wells with two populated subbands, are  investigated. At small magnetic fields the current-induced oscillations are found to be related to Landau-Zener transitions between Landau levels inside the lowest subband. The period of these oscillations is proportional to the magnetic field.  At high magnetic fields a different kind of quantum oscillations are observed. With respect to the dc bias  these resistance oscillations are quasi-periodic with a period that is independent of the magnetic field. At a fixed electric current the oscillations are periodic in inverse magnetic field.  The period is independent of the $dc$ bias. The proposed model considers these oscillations as a result of  joint effect of the Joule heating in the systems with discrete spectrum and the spatial variations of the energy separation  between two subbands, which is  induced by the electric current. Obtained results indicate the feasibility of considerable modification of the electron spectrum by applied electric current in two dimensional electron systems.

\begin{acknowledgements}

S. V. thanks I. L. Aleiner for valuable help with the theoretical model and discussion. Work was supported by  National Science Foundation  (DMR 1104503) and  the Russian Foundation for Basic Research, project no. 11-02-00925.

\end{acknowledgements}

{}

\end{document}